\documentclass[aps,twocolumn,pra,superscriptaddress]{revtex4-1}
\usepackage[english]{babel}
\usepackage[latin1]{inputenc}
\usepackage{hyperref}
\usepackage{amsmath, amssymb, amsfonts}
\usepackage{amssymb,xcolor}
\usepackage{graphicx}
\graphicspath{{images/}}
\usepackage{exscale}
\usepackage{graphicx}
\usepackage{amsmath}
\usepackage{latexsym}
\usepackage{amsfonts}
\usepackage{amssymb}

\DeclareMathOperator{\Tr}{Tr}
\DeclareMathOperator{\Id}{ {\bf 1}}

\def\pmx{\begin{pmatrix}}
\def\emx{\end{pmatrix}}

\newcommand{\ket}[1]{|#1\rangle}
\newcommand{\bra}[1]{ \langle #1 \,  |}

\DeclareMathOperator{\eins}{ {\bf 1}}

\begin{document} 

\title{Scale invariance of entanglement dynamics in Grover's quantum search algorithm}

\author{M. Rossi}\email{matteo.rossi@unipv.it}
\affiliation{Dipartimento di Fisica and INFN-Sezione di Pavia,
Via Bassi 6, 27100 Pavia, Italy}

\author{D. Bru{\ss}}
\affiliation{Institut f{\"u}r Theoretische Physik III, 
Heinrich-Heine-Universit{\"a}t D{\"u}sseldorf, D-40225 D{\"u}sseldorf, Germany}
\author{C. Macchiavello}

\affiliation{Dipartimento di Fisica and INFN-Sezione di Pavia,
Via Bassi 6, 27100 Pavia, Italy}

\begin{abstract}

We calculate the amount of entanglement of the multiqubit quantum states employed in the 
Grover algorithm, by following its dynamics at each step of the computation.  
We show that genuine multipartite entanglement is always present. 
Remarkably, the dynamics of any type of entanglement as well
as of genuine multipartite entanglement 
is independent of the number $n$ of qubits for large $n$, 
thus exhibiting a scale invariance property. We compare this result with the entanglement dynamics induced by a 
fixed-point quantum search algorithm. 
We also investigate criteria for efficient simulatability in
the context of Grover's algorithm.

\end{abstract}

\maketitle

\section{Introduction}

Although it is well-known that entanglement represents an essential ingredient
in quantum communication and information, its role in the speed-up of 
quantum computational processes is not yet fully understood and still 
represents a debated question \cite{jl,gott,vidal,nest}.
In particular, it is of great interest to investigate the 
role of multipartite entanglement in quantum algorithms.
In Shor's algorithm multipartite entanglement was proved 
to be necessary to achieve exponential computational speed-up with quantum 
resources \cite{jl}. Moreover, more recently it was shown that multipartite 
entangled states are employed in the Deutsch-Jozsa algorithm and in the first
step of the Grover algorithm \cite{ent-algo}. 
%A related central issue is the 
%possibility of efficiently simulating quantum computations with classical means.
%In this context a protocol to classically simulate quantum computations that
%involve a limited amount of entanglement for pure states of  $n$ qubits   
% was presented in \cite{vidal}.
In our work we investigate in detail 
%the evolution of 
the entanglement properties
in the Grover algorithm \cite{grover}, namely
we study the behaviour of entanglement of the states of $n$ qubits 
%that are employed in the algorithm 
along the whole computational process, and disclose in 
particular a noteworthy scale invariance property of its dynamics
 in terms of the geometric measure of entanglement (GME). 
Previous works on the entanglement dynamics in Grover's algorithm
considered other entanglement measures and focused only 
on bipartite entanglement (see, for example, \cite{fang,orus}).
We also study the entanglement dynamics in the fixed-point 
$\pi/3$ quantum search algorithm \cite{fixed,fixed2} and show that 
it turns out to be qualitatively similar to the Grover case.
 
This paper is organised as follows. In Sect. \ref{s:g} we consider
the Grover quantum search algorithm and study its entanglement dynamics for
any number of qubits in the cases of one and two solutions to the search
problem. In Sect. \ref{s:fp} we compare these results with the entanglement
dynamics in fixed-point search algorithms. Finally, in Sect. \ref{s:conc}
we summarise the main results and comment on their possible relations to
classical efficient simulatability of Grover's algorithm.

\section{Entanglement dynamics in Grover's algorithm}
\label{s:g}

Let us remind the reader that the Grover search algorithm \cite{grover} employs
pure states of $n$ qubits which are initially prepared in an equally weighted
superposition of all computational basis states $\ket{\psi_0}=\frac{1}{\sqrt{2^n}}\sum_{x=0}^{2^n-1}\ket{x}$, which can be more conveniently written as
\begin{equation}\label{starting}
\ket{\psi_0}=\sqrt{\frac{N-M}{N}}\ket{X_0}+\sqrt{\frac{M}{N}}\ket{X_1},
\end{equation}
where $N=2^n$ and $M$ is the number of searched items (in the following also
referred to as ``solutions'' of the search problem).
Here, 
$\ket{X_1}=\frac{1}{\sqrt{M}}\sum_{x_s}\ket{x_s}$ represents 
the superposition of all the states $\ket{x_s}$ that are solutions (i.e. searched
items), 
and $\ket{X_0}=\frac{1}{\sqrt{N-M}}\sum_{x_n}\ket{x_n}$ denotes the 
superposition of all the states $\ket{x_n}$ that are not searched for.
The global state after $k$ iterations of the Grover operation $G$ has the form
\cite{aharonov,nc}

\begin{equation}\label{iter}
\ket{\psi_k}\equiv G^k\ket{\psi_0}=
\cos \theta_k \ket{X_0}+\sin\theta_k\ket{X_1},
\end{equation}
with $\theta_k=(k+1/2)\theta$ and $\theta=2\sqrt{M/N}$ in the limit $M\ll N$. 
The unitary operation $G$ is usually decomposed in two basic 
blocks, $G={\cal I}U$, where $U$ represents the oracle call, 
i.e. $U=\Id-2\ket{X_1}\bra{X_1}$, and ${\cal I}$ is the inversion about the mean 
operation, namely ${\cal I}=-(\Id-2\ket{\psi_0}\bra{\psi_0})$.
The operation $G$ is repeated until the state  $\ket{\psi_k}$ overlaps as 
much as possible with  $\ket{X_1}$, namely for 
$k_{opt}=CI[(\pi/\theta-1)/2]$, where $CI[x]$ denotes the closest integer to 
$x$. In the limit $M \ll N$, the optimal number of iterations is 
$k_{opt}=CI[\frac{\pi}{4}{\sqrt{N/M}}-\frac{1}{2}]$, i.e. it is proportional 
to the square root of $N$. 
In the following we will consider the condition $M\ll N$ to be always fulfilled.

We will now study the entanglement properties of the states (\ref{iter})
as functions
of the number of iterations $k$ and the number of qubits $n$ for a fixed number
of solutions. We will quantify the amount of entanglement by the GME \cite{wg}, which for a pure $n$-partite state 
$\ket{\psi}$ reads
\begin{equation}\label{geom}
E_q(\ket{\psi})=1-\max_{\ket{\phi}\in S_q}|\bra{\psi}\phi\rangle |^2,
\end{equation}
where $S_q$ is the set of $q$-separable states, namely states that are
separable for $q$ partitions of the $n$-qubit system. 
The GME represents a suitable  
entanglement measure when multi-partite systems are taken into account. 
Notice that $E_n$ quantifies the amount of 
entanglement of any kind contained in the global system, i.e. it is 
non-vanishing even for states showing 
entanglement just between two subsystems, while $E_2$ quantifies
genuine multipartite entanglement \cite{hierarchies}.

\subsection{Single solution}

Let us first consider the case of a single solution to the search problem 
($M=1$). W.l.o.g., as will be proved later, we consider
the state $\ket{X_1}$ representing the solution to be invariant under 
any permutation of the $n$ qubits (e.g. $\ket{111...1}$). Therefore, the state 
$\ket{\psi_{k,M=1}}$ at step $k$ of the algorithm is also permutation invariant for all $k$'s. 
Let us first compute $E_n$ for this set of states for varying $k$.
Due to this symmetry property, the search for the maximum in Eq. \eqref{geom}
can be restricted to  symmetric separable states 
$\ket{\phi}^{\otimes n}$ \cite{symm}, so that the 
maximisation involves only the two parameters $\alpha\in [0,\pi]$ and 
$\beta\in[0,2\pi]$ that define the single qubit state 
$\ket{\phi}=\cos{\frac{\alpha}{2}}\ket{0} + e^{i\beta}\sin{\frac{\alpha}{2}}
\ket{1}$. Furthermore, since $\theta_k \in [0,\pi/2]$ the 
coefficients of $\ket{\psi_k}$ are all positive and the optimal value of the 
phase factor can be fixed to $\beta=0$. 

The GME $E_n$ for a single solution then takes the 
form
\begin{eqnarray}\label{geom_n1}
E_n(\ket{\psi_{k,M=1}})&=&
1-\max_{\alpha}\Big|\frac{\cos\theta_k}{\sqrt{2^n-1}}
\Big[\big(\cos\frac{\alpha}{2}+\sin\frac{\alpha}{2}\big)^n
\nonumber \\
&&-\sin^n\frac{\alpha}{2}\Big]
+\sin\theta_k\sin^n\frac{\alpha}{2}\Big|^2
\end{eqnarray}
The optimal value of $\alpha$ can then be found by setting 
$t=\tan \frac{\alpha}{2}$ and calculating the derivative of the overlap 
explicitly, which reduces to finding the root of a polynomial in $t$.

\begin{figure}[t!]
\centering
\includegraphics[width=1\columnwidth]{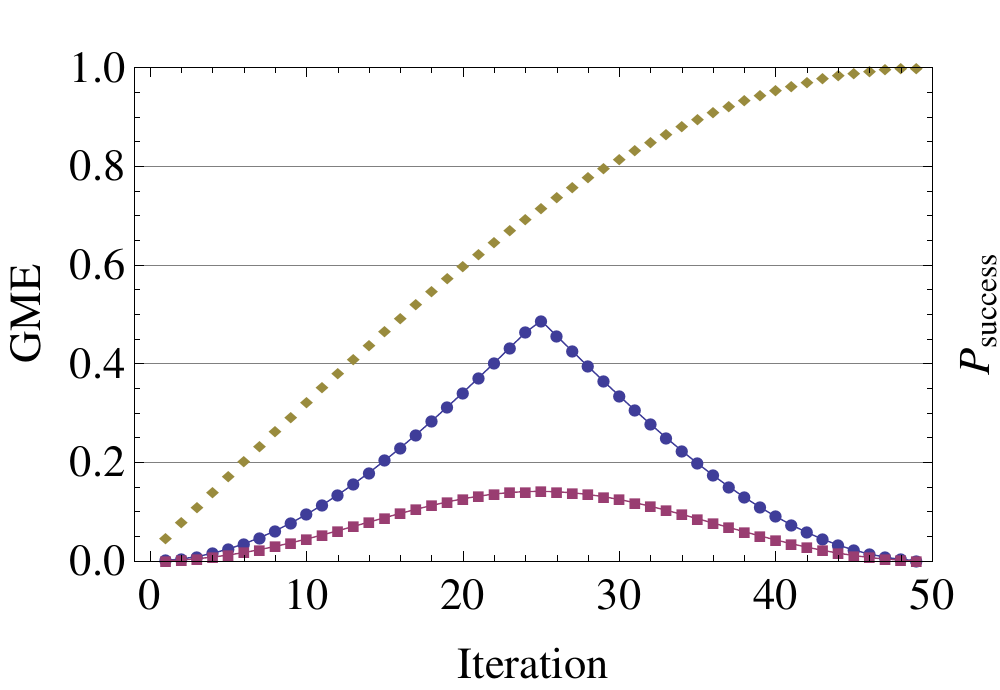}
\caption{(Color online) Evolution of entanglement as 
a function of the number of steps $k$,  for a single searched 
item, with $n=12$ qubits and $k_{opt}=49$.
%The optimal number of iterations is $k_{opt}=49$. 
$E_n(\ket{\psi_{k,M=1}})$
is depicted by blue dots, while 
$E_2(\ket{\psi_{k,M=1}})$ by purple squares. The yellow dots represent the success probability.}\label{evo_1}
\end{figure}

In Fig. \ref{evo_1} we report the behaviour of $E_n(\ket{\psi_{k,M=1}})$ 
for $n=12$: The entanglement increases in the first half of 
iterations, achieves the maximal value of about $1/2$, and then decreases to 
zero as soon as the optimal number of iterations is reached. 
This behaviour is qualitatively similar to the ones shown in \cite{fang,orus}, 
where the dynamics of both the two-qubit concurrence and the Von Neumann entropy of the half-qubit reduced state was studied. 

In order to quantify only genuine multipartite entanglement we will now
calculate $E_2$. The expression of $E_2(\ket{\psi})$ can be rewritten as 
\cite{multi} 
\begin{equation}\label{multi}
E_2(\ket{\psi}) =1-\max_{P}\max_{\mu}\mu^2,
\end{equation}
where the $\mu$'s are the Schmidt coefficients of $\ket{\psi}$ with respect 
to a fixed bipartition $P:Q$, and $\max_P$ denotes the maximisation over all 
possible bipartitions. Notice that, since the regarded state 
$\ket{\psi_{k,M=1}}$ is 
permutation invariant, we need to check only $\lfloor n/2\rfloor$ 
bipartitions, where $\lfloor x\rfloor$ is the largest integer smaller or 
equal to $x$. 
In order to find the maximal Schmidt coefficient of $\ket{\psi_{k,M=1}}$ among all 
possible bipartitions we fix a generic bipartite splitting $P:Q$, where $P$ is composed of $m$ qubits and $Q$ of the remaining $n-m$, and compute the eigenvalues of the reduced density operator $\rho_P=\Tr_Q[\ket{\psi_{k,M=1}}\bra{\psi_{k,M=1}}]$, given by the following 
$2^m\times 2^m$ matrix
\begin{equation}\label{AdagaA}
\rho_P=
\begin{pmatrix}
a&\dots &a &b\\
\vdots &\ddots &\vdots & \vdots\\
a&\dots & a &b\\
b&\dots & b & c\\
\end{pmatrix},
\end{equation}
where $a=2^{n-m}A^2$, $b=a-A(A-B)$, and 
$c=a-A^2+B^2$, with $A=\cos\theta_k/\sqrt{2^n-1}$ and $B=\sin\theta_k$. 
The maximal eigenvalue of the above matrix is given by
\begin{equation}
\lambda_{max}=\frac{1}{2}+\frac{1}{2}\big[1-4(2^m-1)(2^{n-m}-1)
A^2(A-B)^2\big]^{\frac{1}{2}}\;.
\end{equation}
The above expression shows that the bipartition that leads to the maximum
eigenvalue corresponds to $m=1$ for all values of $k$. According to Eq. 
\eqref{multi}, the multipartite GME 
$E_2$ takes the explicit form

\begin{eqnarray}\label{EG2}
E_2(\ket{\psi_{k,M=1}})=\\ \nonumber
\frac{1}{2}-\frac{1}{2}\Big[1
-4&&\frac{2^{n-1}-1}{2^n-1}\cos^2\theta_k\big(\frac{\cos\theta_k}{\sqrt{2^n-1}}-\sin\theta_k\big)^2\Big]^{\frac{1}{2}}\;.
\end{eqnarray}
This result shows that genuine multipartite entanglement has a qualitative 
similar 
behaviour as $E_n(\ket{\psi_{k,M=1}})$ (see Fig. \ref{evo_1}), even if it achieves a 
maximum of about $0.14$ 
%$0.14=\frac{\sqrt 2-1}{2}$
and the curve is derivable in that point. 
Notice also that 
$E_2(\ket{\psi_{k,M=1}})$ is symmetric with respect to $k_{opt}/2$.

We will now show that the entanglement dynamics in the Grover algorithm, namely
the behaviour of $E_n$ and $E_2$ during
the operation of the algorithm, does not depend on the number of qubits $n$, 
thus exhibiting the property of scale invariance. For $2^n\gg 1$ the two 
entanglement measures take the simple forms
\begin{eqnarray}
&E_n(\ket{\psi_{k,M=1}})&\simeq
\begin{cases}
\sin^2\theta_k & \text{for $\theta_k\le \pi/4$},\\
\cos^2\theta_k & \text{for $\theta_k >\pi/4$},
\end{cases} \nonumber \\
&E_2(\ket{\psi_{k,M=1}})&\simeq\frac{1}{2}\Big[1-\big(1-\frac{1}{2}\sin^22\theta_k\big)^{\frac{1}{2}}\Big].
\end{eqnarray}
%E_n(\ket{\psi_{k,M=1}})\simeq
%\begin{cases}
%\sin^2\theta_k & \text{for $\theta_k\le \pi/4$},\\
%\cos^2\theta_k & \text{for $\theta_k >\pi/4$},
%\end{cases}
%\end{equation}
%and
%
%\begin{equation}
%E_2(\ket{\psi_{k,M=1}})\simeq\frac{1}{2}\Big[1-\big(1-\frac{1}{2}\sin^2 
%2\theta_k\big)^{\frac{1}{2}}\Big]\;.
%\end{equation}
Both expressions depend only on 
$\theta_k\simeq\frac{\pi}{2}k/k_{opt}$, namely on 
$k/k_{opt}$, and not on $k$ and $n$ separately. Therefore, the entanglement 
dynamics of the Grover algorithm is scale invariant in the sense that it only depends on
the number of steps taken, relative to the total number, but not on the length of the list.
%This statement holds  for $2^n\gg 1$, and numerical simulations show that this limit is achieved for $n\geq 15$.

We want to point out that all the results presented so far, even if they were 
explicitly derived for permutation invariant states, hold for any instance of 
the Grover algorithm with one searched item, i.e. $M=1$. 
The number of possible single searched items in the Grover algorithm 
is $2^n$, which corresponds to the number of distinct states $\ket{X_1}$.
All of these states can be achieved from a symmetric one 
by applying tensor products of $\sigma_x$ Pauli operators and identity 
operators $\eins$ (e.g. $\ket{001...1}=\sigma_{x1}\otimes\sigma_{x2}\otimes
\eins_3....\ket{111...1}$). 
Since these operations are local, 
they do not change the entanglement content of the resulting state.

\subsection{Two solutions}

\begin{figure}[t!]
\centering
\includegraphics[width=1\columnwidth]{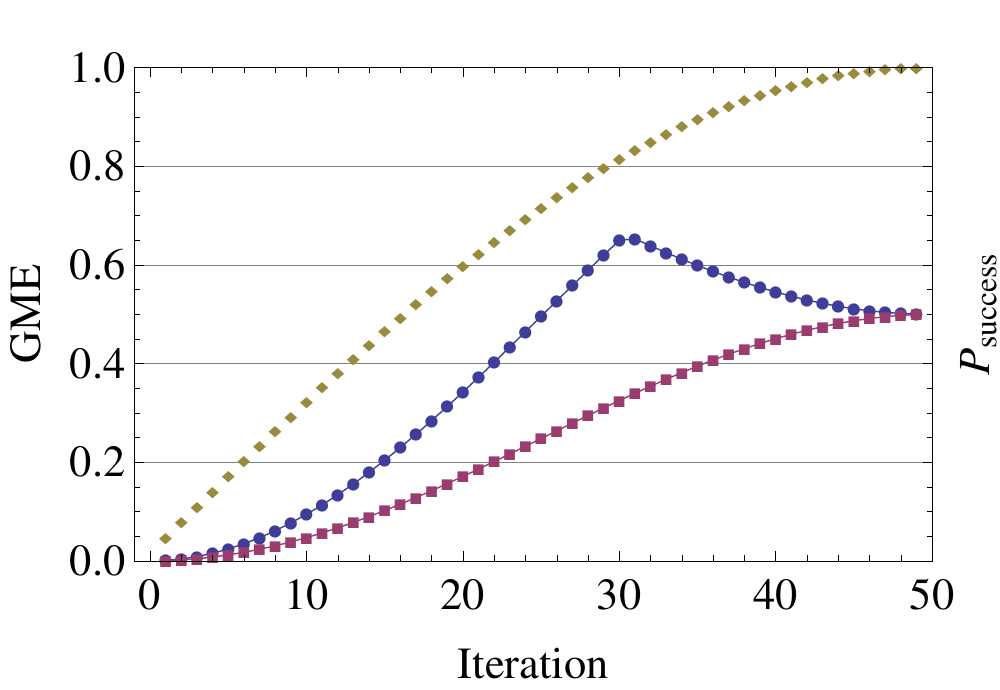}
\caption{(Color online) The GME 
as a function of the
number of steps $k$, for two symmetric solutions of the search 
problem. The number of qubits is $n=13$, and %the optimal number of iterations is
$k_{opt}=49$. 
$E_n(\ket{\psi_{k,M=2}})$ is given by blue dots, 
$E_2(\ket{\psi_{k,M=2}})$ by purple 
squares. The yellow dots represent the success probability.}\label{evo_2}
\end{figure}

Let us now consider the case of two searched items (i.e. $M=2$). 
As an illustrative example we will consider the case in which both 
$\ket{00\dots0}$ and $\ket{11\dots1}$ are solutions of the search problem, 
thus the state $\ket{X_1}$ is a GHZ state composed of $n$ qubits, and the 
state at each step of the computation is permutation invariant. 
The measure of any entanglement $E_n$ is given by

\begin{eqnarray}
&&E_n(\ket{\psi_{k,M=2}})=1-\max_\alpha\Big|\frac{\cos\theta_k}{\sqrt{2^n-2}}
\Big[\big(\cos\frac{\alpha}{2}+\sin\frac{\alpha}{2}\big)^n \nonumber\\
&& 
-\Big(\cos^n\frac{\alpha}{2}+\sin^n\frac{\alpha}{2}\Big)\Big]
+\frac{\sin\theta_k}{\sqrt 2}\Big(\cos^n\frac{\alpha}{2}
+\sin^n\frac{\alpha}{2}\Big)\Big|^2,\nonumber 
\end{eqnarray}
We maximised this quantity numerically; in  Fig. \ref{evo_2} we show the behavior 
for $n=13$.
Notice that after $k_{opt}$ iterations, the measure $E_n(\ket{\psi_{k,M=2}})$ is no longer 
zero but equal to $1/2$. That is because the final state is no longer fully 
separable but instead it is close to the GHZ state. In this case 
the maximal value that the entanglement reaches during 
the computation is about $2/3$, i.e. higher than the case $M=1$. Furthermore, 
this maximum is no longer reached at half of the optimal number of steps
$k_{opt}$, but in a later 
step, i.e. $k/k_{opt}\simeq 0.61$. 

Regarding genuine multipartite entanglement, $E_2$ with two 
symmetric solutions can be computed by following an analogous procedure as for a 
single solution. The reduced density matrix for the general bipartite 
splitting $P:Q$, where $m$ qubits are in  $P$ and $n-m$ in $Q$, is now 
given by
\begin{equation}\label{AdagaA2}
\rho_P=
\begin{pmatrix}
c &b& \dots & b&d\\
b&a &\dots &a &b\\
\vdots &\vdots &\ddots &\vdots & \vdots\\
b&a&\dots &a &b\\
d& b&\dots &b&c\\
\end{pmatrix},
\end{equation}
where $a,b,c$ and $A,B$ are given below Eq. (\ref{AdagaA}), 
and $d=a-2A(A-B)$.
It turns out that again the maximum eigenvalue corresponds to 
the bipartite splitting with $m=1$, and
$E_2(\ket{\psi_{k,M=2}})$ can be expressed analytically as
\begin{equation}\label{EG2bis}
E_2(\ket{\psi_{k,M=2}})=1-\frac{2^n-4}{2^n-2}\cos^2\theta_k
-\Big(\frac{\cos\theta_k}{\sqrt{2^n-2}}+\frac{\sin\theta_k}{\sqrt 2}\Big)^2.
\end{equation}
This result is shown in Fig. \ref{evo_2}. Notice that multipartite 
entanglement has a different behaviour from $E_n(\ket{\psi_{k,M=2}})$. 
It is a monotonically increasing function that approaches the maximum value 
of $1/2$ when the computation stops. 

In the asymptotic limit $2^n\gg 1$ the GME 
can be
expressed  as
\begin{eqnarray}
&E_n(\ket{\psi_{k,M=2}})&\simeq
\begin{cases}
\sin^2\theta_k & \text{for $\theta_k\le \arccos 1/\sqrt 3$},\\
\frac{1+\cos^2\theta_k}{2} & \text{for $\theta_k >\arccos 1/\sqrt 3$}\;,
\end{cases}\nonumber \\
&E_2(\ket{\psi_{k,M=2}})&\simeq\frac{1}{2}\sin^2\theta_k\;.
\end{eqnarray}
As a consequence, both quantities exhibit the same scale invariance 
behaviour as discussed above for the case with one searched item, i.e. $M=1$.
We point out that the above results  can be generalized 
to those search problems in which the two solutions are different 
in all digits, but not to all search  problems with $M=2$.

\section{Entanglement dynamics in the fixed-point $\pi/3$ quantum search 
algorithm}
\label{s:fp}

In the previous section we have shown
that  the amount of entanglement is non-vanishing during 
the Grover algorithm and that its behaviour is scale invariant 
for a single solution to the search problem and in some instances of two 
solutions.
We will now show that a similar entanglement dynamics can be found in the 
fixed-point $\pi/3$ quantum search. 
This kind of quantum search algorithm was first introduced in \cite{fixed} 
to overcome the fact that the Grover algorithm might lead to a high error 
probability if the number of solutions $M$ is unknown, since it
requires to stop at a precise iteration $k_{opt}$, which depends on $M$.
In contrast the $\pi/3$ quantum search always converges to the solutions, 
and thus it can be regarded as a fixed-point algorithm, even if it is never 
as fast as the standard Grover algorithm.
\begin{figure}[t!]
\centering
\includegraphics[width=1\columnwidth]{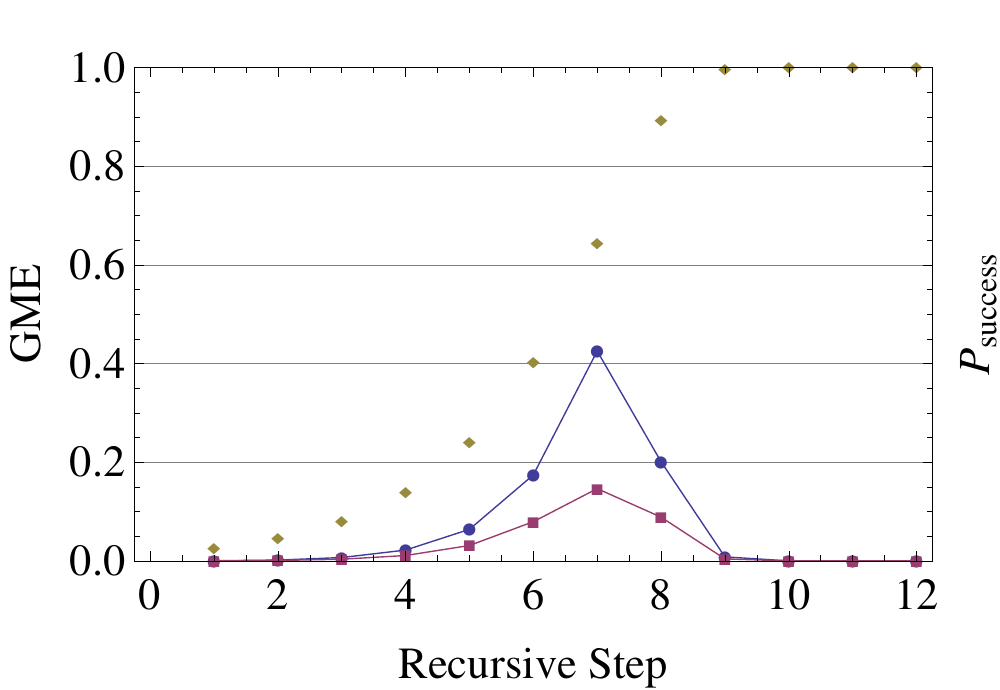}
\caption{(Color online) Evolution of entanglement in the $\pi/3$ search as 
a function of the recursive step $m$,  for a single searched 
item, with $n=12$ qubits. Here and below in Fig. \ref{evo_2_pi3} blue dots,
purple squares and yellow dots denote $E_n$, $E_2$ and the success probability, respectively.
%The optimal number of 
%iterations is $k_{opt}=49$. Any kind of entanglement, i.e. $E_n(\ket{\psi_{k,M=1}})$,
%is represented by blue dots, while genuine multipartite entanglement
%$E_2(\ket{\psi_{k,M=1}})$ is depicted by purple squares.
}\label{evo_1_pi3}
\end{figure}

A possible way to realise such a fixed-point search is to slightly modify 
the operations $U$ and ${\cal I}$ in order to produce a $\pi/3$ shift instead 
of a $\pi$ shift \cite{fixed}, i.e.
\begin{align}
&U\longrightarrow U_{\frac{\pi}{3}}=\Id-(1-e^{i\frac{\pi}{3}})\ket{X_1}\bra{X_1},\\ \nonumber
&{\cal I}\longrightarrow {\cal I}_{\frac{\pi}{3}}=-(\Id-(1-e^{i\frac{\pi}{3}})\ket{\psi_0}\bra{\psi_0}).
\end{align}
Then, the sequence of gates to be applied is defined by the 
following recursive formula
\begin{align}\label{recursion}
& A_{m+1}=A_m {\cal I}_{\frac{\pi}{3}} A_m^\dagger U_{\frac{\pi}{3}} A_m,\\ \nonumber
& A_0=\Id.
\end{align}

\begin{figure}[t!]
\centering
\includegraphics[width=1\columnwidth]{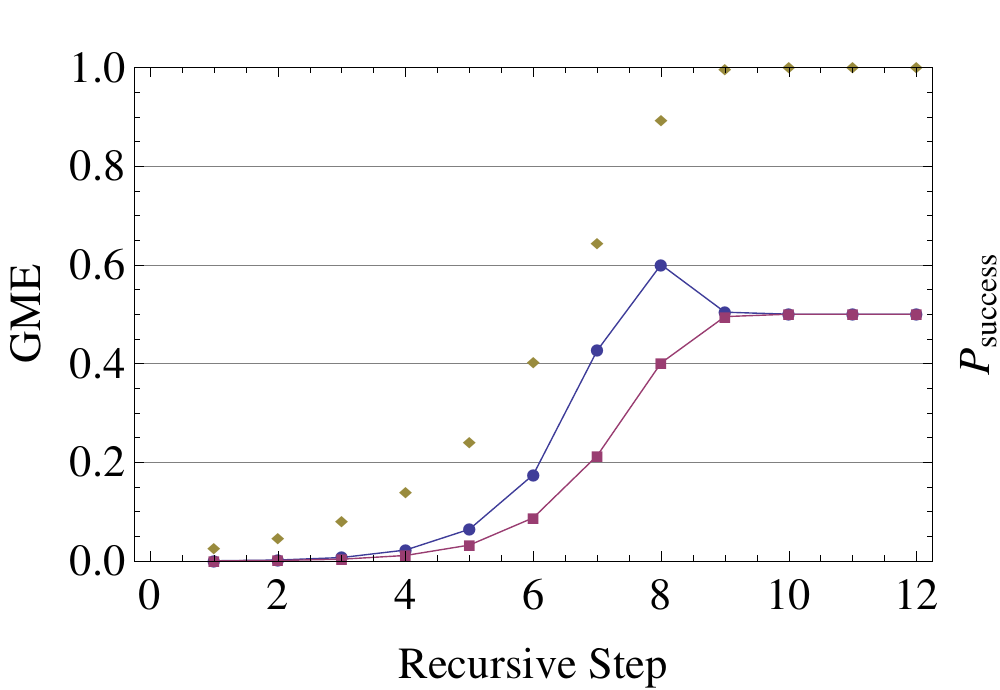}
\caption{(Color online) Evolution of entanglement in the $\pi/3$ search as 
a function of the recursive step $m$,  for two symmetric solutions of the search problem, with $n=13$ qubits.
%The optimal number of 
%iterations is $k_{opt}=49$. Any kind of entanglement, i.e. $E_n(\ket{\psi_{k,M=1}})$,
%is represented by blue dots, while genuine multipartite entanglement
%$E_2(\ket{\psi_{k,M=1}})$ is depicted by purple squares.
}\label{evo_2_pi3}
\end{figure}

We now compute both $E_n$ and $E_2$ for the employed states at each recursive 
step $m$ of the evolution. The results were obtained  numerically and 
are shown in Figs. 
\ref{evo_1_pi3} and \ref{evo_2_pi3} for both one and two solutions.
Notice that the entanglement behaviour of both $E_n$ and $E_2$ is similar to 
the dynamics of the standard Grover algorithm. These results indicate 
that entanglement plays the same crucial role in both algorithms, 
even if a scale invariance property cannot be proved in the 
fixed-point algorithm case.

%We have thus shown analytically that  the amount of multipartite
%entanglement is non-vanishing during the Grover algorithm, and that 
%its characteristic behaviour is scale invariant, i.e. independent of the length of the
%list to be searched.

\section{Conclusions}
\label{s:conc}

In summary, we have studied the evolution of entanglement in Grover's 
algorithm (for a small number of searched items),  quantifying it via  
the GME. 
In particular, we give an explicit formula for  the amount of genuine 
multipartite entanglement, which is proven to be always non-zero throughout 
the computation. Interestingly, the dynamics of entanglement shows
the behaviour of scale  invariance, i.e. counter-intuitively 
the amount of entanglement employed in the algorithm
does not depend on the length of the searched list, but
only on the number of steps taken, relative to the optimal number
of steps.  Since scale invariance is an important phenomenon in several areas of
physics and mathematics, our results may open new avenues in the understanding of scale 
invariance properties of entanglement in other contexts, such as 
for example in many-body systems and phase 
transitions.
%As our  measure for entanglement
%corresponds to a distance in Hilbert space,
%our observation may hint at some self-similar structure of  Hilbert
%space (i.e. certain distances of  entangled states 
%remain unchanged when zooming into a smaller-dimensional subspace)
%- we will leave this conjecture for 
%further investigation. 
We have also compared the Grover search entanglement dynamics with the one of a different kind of search algorithm, i.e. the $\pi/3$ quantum 
search, and we have showed that they exhibit a similar behavior.

{As a final comment, we may wonder whether the presence of true 
multipartite 
entanglement means that Grover's algorithm cannot be simulated efficiently
by classical means.}
By efficient classical simulation of Grover's algorithm we mean that, given a database of $n$ qubits, i.e. $2^n$ items, 
it is possible to classically simulate Grover's algorithm 
with a total cost that scales as $\sqrt{2^n}\text{poly}(n)$.
We will now show that well-known criteria which guarantee efficient
simulatability do not apply for Grover's algorithm. 
According to the Gottesman-Knill theorem  \cite{nc,gott},  
if a quantum computation starts in a computational basis state and 
involves only stabilizer gates (i.e. Hadamard, CNOT, phase gates and 
measurement of operators in the Pauli group), then it can be efficiently 
simulated on a classical computer. However, it can be easily shown that
${\cal I}$ transforms an element belonging to the Pauli group, i.e. 
$\sigma_{z}\otimes \eins^{(n-1)}$, to an operator that no longer belongs to 
the Pauli group,
and therefore it cannot be implemented by stabilizer gates.
Let us also consider the simulatability criterion introduced in 
\cite{vidal}, based on the maximal Schmidt rank $\chi$ of $\ket{\psi}$ over 
all possible bipartitions. According to \cite{vidal}, if $\chi$ does not 
exceed  $\text{poly}(n)$ in a computation that consists of  $\text{poly}(n)$ 
elementary gates (i.e. one- and two-qubit gates),
then the computation  can be classically simulated efficiently.
We notice that for states of the form (\ref{iter}), $\chi$ is upper bounded 
by $M+1$. However, although there exists a decomposition of the Grover 
operation $G$ into
$\text{poly}(n)$ elementary gates \cite{implem,nc}, the state after 
the action of each two-qubit gate does not have a simple symmetric form and we 
no longer can keep track of the maximal Schmidt rank. 
Therefore we cannot conclude efficient simulatability. 

The above results show that, although the Grover operation cannot be 
implemented 
by stabilizer gates and therefore the Knill-Gottesman theorem cannot be 
applied, the employed states at each Grover iteration are only slightly 
entangled according to the criterion suggested in \cite{vidal}. 
These insights are  nevertheless  not sufficient to answer the question of
 simulatability of the Grover algorithm, which  at present remains open.

\section{Acknowledgements}

This work was financially supported by DFG. 
%(Deutsche Forschungsgemeinschaft).
MR is supported by DAAD 
%(Deutscher Akademischer Austausch-Dienst)
and acknowledges the hospitality of the Heinrich-Heine Universit{\"a}t D\"usseldorf.

\end{document}